\documentclass[aps,amsfonts,amsmath,prd,preprint,nofootinbib]{revtex4-1}
\pdfoutput=1
\usepackage{graphicx}

\newcommand{\beq}{\begin{equation}}
\newcommand{\eeq}{\end{equation}}

\def\({\left(}
\def\){\right)}

\begin{document}

\title{Instability of an emergent universe}

\author{Audrey T. Mithani and Alexander Vilenkin}
\address{
Institute of Cosmology, Department of Physics and Astronomy,\\ 
Tufts University, Medford, MA 02155, USA}

\begin{abstract}

Oscillating solutions to the effective equations of Loop Quantum Cosmology have been suggested for the role of an `eternal seed', providing a possible starting point for the emergent universe scenario.  We investigate the stability of a particular model, sourced by a homogeneous massless scalar field and a negative cosmological constant, with respect to small perturbations and to quantum collapse.  We find that the model has
perturbatively stable and unstable solutions, with both types of solutions occupying significant regions of the parameter space.  All solutions are unstable with respect to collapse by quantum tunneling to zero size.  We discuss the possibility that the state resulting from the collapse is non-singular, so it may tunnel back to the oscillating regime. We argue that the universe is then likely to evolve to states of very large size with large particle occupation numbers.

\end{abstract}

\maketitle

\section{Introduction}

Cosmic inflation may be eternal to the future, but it cannot be eternal to the past.
More specifically, if the average expansion rate $H_{av}$ along some geodesic is positive, that geodesic must be past-incomplete \cite{BGV}.  This suggests that the inflationary phase of our universe had some sort of a beginning, perhaps a quantum nucleation from `nothing' \cite{AV82}.

A possible alternative is a universe which existed as an eternal ``seed'' before sprouting into our inflating universe.  In Refs. \cite{EM2004} and \cite{Murugan}, Ellis and collaborators proposed a mechanism for this type of ``emergent'' universe scenario.  Their model relies on a scalar field in a potential which is asymptotically flat as $\phi \to -\infty$, then reaches a hill as $\phi \to 0$, signaling the beginning of inflation.  So long as the condition $H_{av} = 0$ is satisfied prior to the beginning of inflation, the conclusions of \cite{BGV} do not apply. 

However, for the emergent universe to be a viable scenario, it must be possible to construct a model which is stable for eternity.  One candidate model was discussed by Graham et al \cite{Graham}, who found that a spherical ($K=+1$) universe containing matter with equation of state $w=-2/3$ and a negative cosmological constant $\Lambda <0$ oscillates as a simple harmonic oscillator. The negative $\Lambda$ causes an expanding universe to turn around and recollapse, while the matter component with $w=-2/3$ causes a collapsing universe to bounce.  The result is oscillation. 
A perfect fluid with $w=-2/3$ would be unstable with respect to high-frequency perturbations, but Graham et al argue that a frustrated network of domain walls would have an effective equation of state $w=-2/3$ and be perturbatively stable.  However, it has been shown in \cite{MV} that this `simple harmonic universe' can tunnel from the bounce point to nothing, signaling a quantum instability and indicating that a universe of this type cannot last forever.

A different class of models for an emergent universe are motivated by Loop Quantum Cosmology (LQC), which provides a nonsingular bounce scenario due to a modification in the effective Friedmann equation.  Mulryne et al \cite{Mulryne2005} have shown that such a model has eternally oscillating solutions with positive spatial curvature ($K=+1$), and that inflation may arise with an appropriate potential for the scalar field.  An LQC model with flat spatial geometry ($K=0$) and the energy density given by a massless scalar field $\phi$ and a negative cosmological constant was recently discussed by Mielczarek et al \cite{MSS}, who found a simple oscillating solution in this case.  It should be noted that Loop Quantum Gravity, on which LQC is based, is still an incomplete theory; in this sense the foundations of LQC are not very reliable.  However, LQC has now been studied in great detail and is an interesting (albeit restricted) theory in its own right (see Ref.~\cite{AS} for an up to date review).  It resolves the singularities of FRW models and provides a useful framework for investigating Planck-scale physics.
   
In the present paper, we shall investigate the stability of the oscillating universe of Ref.~\cite{MSS} with respect to classical scalar field perturbations and to quantum tunneling.  In the next Section we introduce the model and its oscillating solution.  The stability of this solution with respect to small perturbations is analyzed in Section III.  We find that unstable modes are present for a significant part of the parameter space of the model, but there are also significant ranges of the parameters where the model is perturbatively stable.  This situation is similar to that found for the simple harmonic universe in Ref.~\cite{Graham}, but there are also some significant differences, as we shall see.

The stability of the oscillating LQC model with respect to quantum tunneling will be investigated in Section IV.  We earlier found in Ref.~\cite{MV2} that in order to stabilize an oscillating universe against quantum collapse in standard Einstein gravity, one would need a matter component with a negative energy density that grows faster than $a^{-6}$ at $a\to 0$, where $a$ is the scale factor.  As we shall discuss in the following Section, the effective Friedmann equation of LQC replaces the matter density $\rho$ with ${\tilde\rho} =\rho(1-\rho/\rho_{max})$, where $\rho_{max}=const$.  The energy density of a massless scalar field is $\rho\propto a^{-6}$, and thus ${\tilde\rho}(a\to 0)\propto - a^{-12}$, suggesting the possibility that the model could be non-perturbatively stable.  We find, however, that the results of Ref.~\cite{MV2} cannot be applied here, because the gravitational action in LQC is significantly different from that in Einstein gravity.   With this difference taken into account, a semiclassical analysis shows that the oscillating LQC solutions are still unstable with respect to quantum collapse.  Our conclusions are summarized and discussed in Section V.

\section{Classical Solutions}

We consider a spatially flat but closed universe, which has a toroidal topology,
\beq
ds^2 = N^2(t) dt^2 - a^2(t) d{\bf x}^2,
\eeq
where $a(t)$ is the scale factor and $N(t)$ is the lapse function.  We shall use the proper time gauge $N=1$ throughout the paper.  Here, the Cartesian coordinates $x^i$ vary in the range
\beq
0\leq x^i\leq L,
\label{periodic}
\eeq
where $L={\rm const}$ and the points $x^i =0$ and $x^i =L$ are identified.  The dynamics of this model in 
LQC is accurately described by the effective Hamiltonian \cite{AS}\footnote{The effective Hamiltonian commonly used in the LQC literature is given by (\ref{Heff}) with $\cos(\ell\beta)$ replaced by $\sin(\ell\beta)$.  These two forms of ${\cal H}$ are related by a simple change of variable $\beta\to\beta + \pi/2\ell$.} 
\beq
{\cal H} = -\frac{3}{8\pi G \gamma^2} \frac{\cos^2(\ell \beta)}{\ell^2} V +{\cal H}_{matter},
\label{Heff}
\eeq
where $V = a^3 L^3$ is the volume, 
\beq
\beta=4\pi G\gamma p_V,
\eeq 
and $p_V$ is the canonical momentum conjugate to $V$.  Here, $\gamma$ is the Barbero-Immirzi parameter, and $\ell^2 = 4\pi \sqrt{3} \gamma G$.  It is usually assumed that $\gamma\sim 1$; then $\ell$ is comparable to the Planck length.  

We consider a model where the energy density is due to a massless scalar field $\phi$ and a negative cosmological constant, $\Lambda <0$.  The corresponding matter Hamiltonian is
\beq
{\cal H}_{matter} = \frac{p_\phi^2}{2V}+\Lambda V \equiv \rho V,
\label{Hmatt}
\eeq
where $p_\phi$ is the momentum conjugate to $\phi$, and we have defined the matter energy density $\rho$.
The classical equations of motion for the canonical variables are then
\begin{eqnarray}
\dot{p}_{\phi} & = & 0 \label{dotp}\\
\dot{\phi} & = & \frac{p_{\phi}}{V} \label{dotphi}\\
\dot{V} & = & -\frac{3}{\gamma}V\frac{\cos(\ell \beta)}{\ell} \sin(\ell \beta) \\
\dot{\beta} & = & -4\pi G \gamma \frac{p_{\phi}^2}{V^2} \label{dotb}.
\end{eqnarray}
where in the final relation we have used the Hamiltonian constraint 
\beq
{\cal H} = 0.
\label{Hconstraint}
\eeq

Squaring the equation for $\dot{V}$ and combining with Eqs.~(\ref{Hmatt}) and (\ref{Hconstraint}), we find the modified Friedmann equation:
\beq
\frac{\dot{a}^2}{a^2} = \frac{\dot{V}^2}{9V^2} = \frac{8\pi G}{3}\rho \(1-\frac{\rho}{\rho_{max}}\),
\label{Friedmann}
\eeq
where 
\beq
\rho_{max}=\frac{\sqrt{3}}{32\pi^2 \gamma^3 G^2}
\eeq
is the maximum energy density that can be attained in this model.\footnote{The same form of the modified Friedmann equation is obtained in the Shtanov-Sahni braneworld model with a timelike extra dimension \cite{SS,Copeland2005}.}  We shall assume that $|\Lambda| \lesssim \rho_{max}$.

The equations of motion for $\phi$ and $p_\phi$ in Eq.~(\ref{dotp}-\ref{dotphi}) result in the familiar wave equation for a homogeneous massless scalar field,
\beq
\ddot{\phi} + 3\frac{\dot a}{a} \dot{\phi} = 0
\eeq
Using Eq.~(\ref{Hmatt}), the matter energy density can be expressed as 
\beq
\rho = \Lambda + \frac{C}{V^2},
\label{rho}
\eeq
where $C=p_\phi^2 /2 = {\rm const}$.

With this matter density, the Friedmann equation (\ref{Friedmann}) has an oscillating solution \cite{MSS}:
\begin{eqnarray}
\label{at}
V(t) & = & V_{min} (2\lambda)^{-1/2} \left( -\cos(\omega t) +1 + 2\lambda
\right)^{1/2} ,\\
\label{omega}
\omega & = & \sqrt{96\pi G \rho_{max} \lambda\( 1+ \lambda \)},
\end{eqnarray}
where $\lambda=|\Lambda|/\rho_{max}$ and 
\beq
V_{min} = \( \frac{C}{\rho_{max} + |\Lambda|} \)^{1/2} \equiv L^3 a_{min}^3.  
\eeq
The universe oscillates at frequency $\omega$ between minimum volume $V_{min}$, where $\rho=\rho_{max}$, and maximum $V_{max} = V_{min}\frac{1+\lambda}{\lambda}$, where $\rho=0$.  

The value of the parameter $L$ depends on the normalization of the scale factor $a$.  We shall choose it so that $a_{min}=1$.  Then $V_{min}=L^3$ and $L = (C/\rho_{max})^{1/6}(1+\lambda)^{-1/6}$.

We may also find a solution for the momentum $\beta$:
\beq
\beta(t) = \left(\frac{8\pi G\gamma^2 \rho_{max}}{3}\right)^{1/2}
\arctan \( \sqrt{\frac{1+\lambda}{\lambda}}\tan \( \frac{\omega t}{2} \) \).
\label{beta}
\eeq

We note that in the limit $\Lambda\to 0$ the solution (\ref{at}) goes into
\beq
V(t) = V_{min}^{(0)} \left( 1 + 24\pi G\rho_{max} t^2 \right)^{1/2},
\label{Vbounce}
\eeq
where $V_{min}^{(0)} = \( {C}/{\rho_{max}} \)^{1/2}$.
This solution, which describes a contracting, bouncing and re-expanding universe,  
 has been discussed earlier by a number of authors \cite{sLQC}.

\section{Perturbative stability}

We generally expect to find particle production in an oscillating universe.  In our model, this would manifest itself in unstable (growing) inhomogeneous fluctuations of the scalar field or the metric.  We shall now check for this instability.

The dynamics of perturbations in LQC has been studied by Agullo et al \cite{Agullo}, with the conclusion that it is accurately given by the usual quantum field theory on the FRW background described by the solutions to the effective field equations.\footnote{It is shown in Ref.~\cite{Param} that this prescription can become inaccurate in the regime where $C\ll 10^2 \ell^6 \rho_{max}$, so that the bounce is at a near-Planckian volume $V_{min}\ll 10 \ell^3$.  Here we shall assume that $C\gg 10^2 \ell^6 \rho_{max}$.}  In our case, the scalar field perturbations should then satisfy the Klein-Gordon equation
\beq
\Box \phi = {\ddot\phi} + 3\frac{\dot a}{a} {\dot\phi} -\frac{1}{a^2} {\bf\nabla}^2 \phi = 0
\label{KG}
\eeq
in the oscillating background (\ref{at}).  Metric perturbations, describing gravitons, satisfy the same equation, so it will be sufficient to investigate the stability of the solutions of Eq.~(\ref{KG}).


It will be convenient to introduce a new variable $\tau=\omega t$.  The field $\phi$ can then be expanded into plane waves, 
\beq
\phi({\bf{x}},\tau) = \sum_{\bf k} \phi_k(\tau) e^{i{\bf{k x}}} ,
\eeq
where the mode functions $\phi_k$ satisfy 
\beq
{\phi''}_k + 3 \frac{a'}{a} {\phi'}_k +\frac{k^2}{\omega^2 a^2}\phi_k = 0.
\label{phieq}
\eeq
Here, primes stand for derivatives with respect to $\tau$.

Periodic boundary conditions in the range (\ref{periodic}) require that  $k_i = 2\pi n_i /L$,
where $n_i$ are integers.  The eigenvalues of the Laplacian are then given by
\beq
k^2 = (2\pi/L)^2 n^2,
\label{k2}
\eeq
where $n^2\equiv n_1^2+n_2^2+n_3^2$.

The scale factor corresponding to the solution (\ref{at}) can be written as
\begin{eqnarray}
\label{at'}
a(\tau)  =  (2\lambda)^{-1/6} \left( -\cos\tau +1 +2\lambda \right)^{1/6} , 
\end{eqnarray}
where we have normalized to $a_{min}=1$.  With this form of the scale factor, the mode equation (\ref{phieq}) becomes
\beq
{\phi''}_k + \frac{\sin{\tau}}{2f(\tau)}
{\phi'}_k +\frac{q^2}{f^{1/3}(\tau)}\phi_k = 0 ,
\label{phikeq}
\eeq
where 
\beq
f(\tau)=-\cos \tau +1+2\lambda
\eeq
and
\beq
q^2 = \frac{(2\lambda)^{1/3} k^2}{\omega^2} =\frac{\pi}{12} \frac{n^2}{GL^2 \rho_{max} (2\lambda)^{2/3} (1+\lambda)}.
\label{q}
\eeq

Eq.~(\ref{phikeq}) is a form of Hill's equation, which is a more general form of the familiar Mathieu equation.  As with the Mathieu equation, we can find stable and unstable regions in the parameter space.  The modes $\phi_k$ represent excitations of the field at a certain momentum ${\bf k}$, so growing solutions to the mode equation signal particle production in that state.

With a transformation to conformal time $d\eta = a^{-1} dt$ and defining $y(\eta) = a(\eta)\phi(\eta)$, Eq.~(\ref{phikeq}) can be brought to the form
\beq
y'' + (k^2 + a''/a)y = 0, 
\label{conformal}
\eeq
where primes stand for derivatives with respect to $\eta$.  One can then use the standard stability analysis using Floquet theory \cite{Magnus}.  We did not follow this path because we could not obtain an analytic solution for $a(\eta)$ and using a numerical solution would take an excessive amount of computer time.  We therefore modified the Floquet method as described in Appendix A, so that it can be directly applied to Eq.~(\ref{phikeq}).
The resulting stability diagram in the parameter space of $q$ and $\lambda$ is shown in Fig.~1.
\begin{figure}[t]
\begin{center}
\includegraphics[width=10cm]{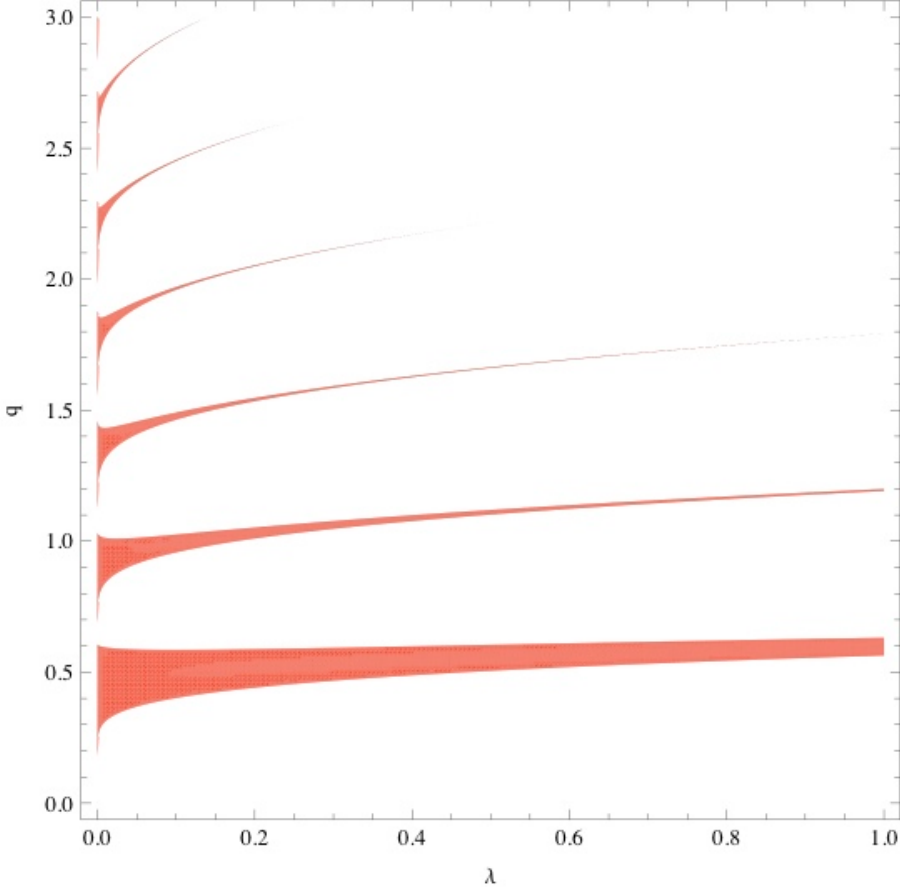}
\end{center}
\label{diagram}
\caption{Stability diagram in the parameter space of $q$ and $\lambda$.  Red bands are unstable.}  
\end{figure}

The parameter $\lambda$ in our model determines the ratio $a_{max}/a_{min}$.  For $\lambda \sim 1$, we have $a_{max}\sim a_{min}$, so the scale factor oscillates about the value $a\sim 1$ with an amplitude also $\sim 1$.  In this regime, the diagram exhibits the characteristic pattern of narrow parametric resonance, with instability confined to narrow bands.  For $\lambda\ll 1$, $a_{max}/ a_{min}\sim \lambda^{-1/3} \gg 1$, so the size of the universe changes by many orders of magnitude in the course of one oscillation.   The instability bands become broader in this limit.

These results can be qualitatively understood as follows.  In the absence of oscillations, the energy spectrum of a scalar field in a compact universe is discrete.  If the oscillation amplitude is relatively small, the oscillations act as a periodic perturbation of frequency $\omega$, and particle production occurs only if $\omega$ is very close to one of the resonant frequencies of the modes.  As the oscillation amplitude gets large, the perturbation effectively includes a wide range of frequencies, so a larger number of modes are affected. 

As unstable mode functions oscillate, their amplitudes grow exponentially, 
\beq
\phi_k(\tau) \propto e^{\alpha \tau} ,
\eeq
with the rate of growth $\alpha$ getting higher as $\lambda$ gets smaller and the instability band widens.  The time evolution of the energy density,
\beq
\rho_k(\tau) = \frac{\omega^2}{2} \left({{\phi'}_k}^2 + \frac{k^2}{\omega^2 a^2} {\phi_k}^2 \right) ,
\eeq
is shown in Fig.~2 for $q=0.55$ and $\lambda = 0.5,~ 0.05,~ 0.01$.  The corresponding growth rates are $\alpha = 0.09,~ 0.23,~ 0.31$, respectively.

\begin{figure}[t]
\begin{center}
\includegraphics[width=10cm]{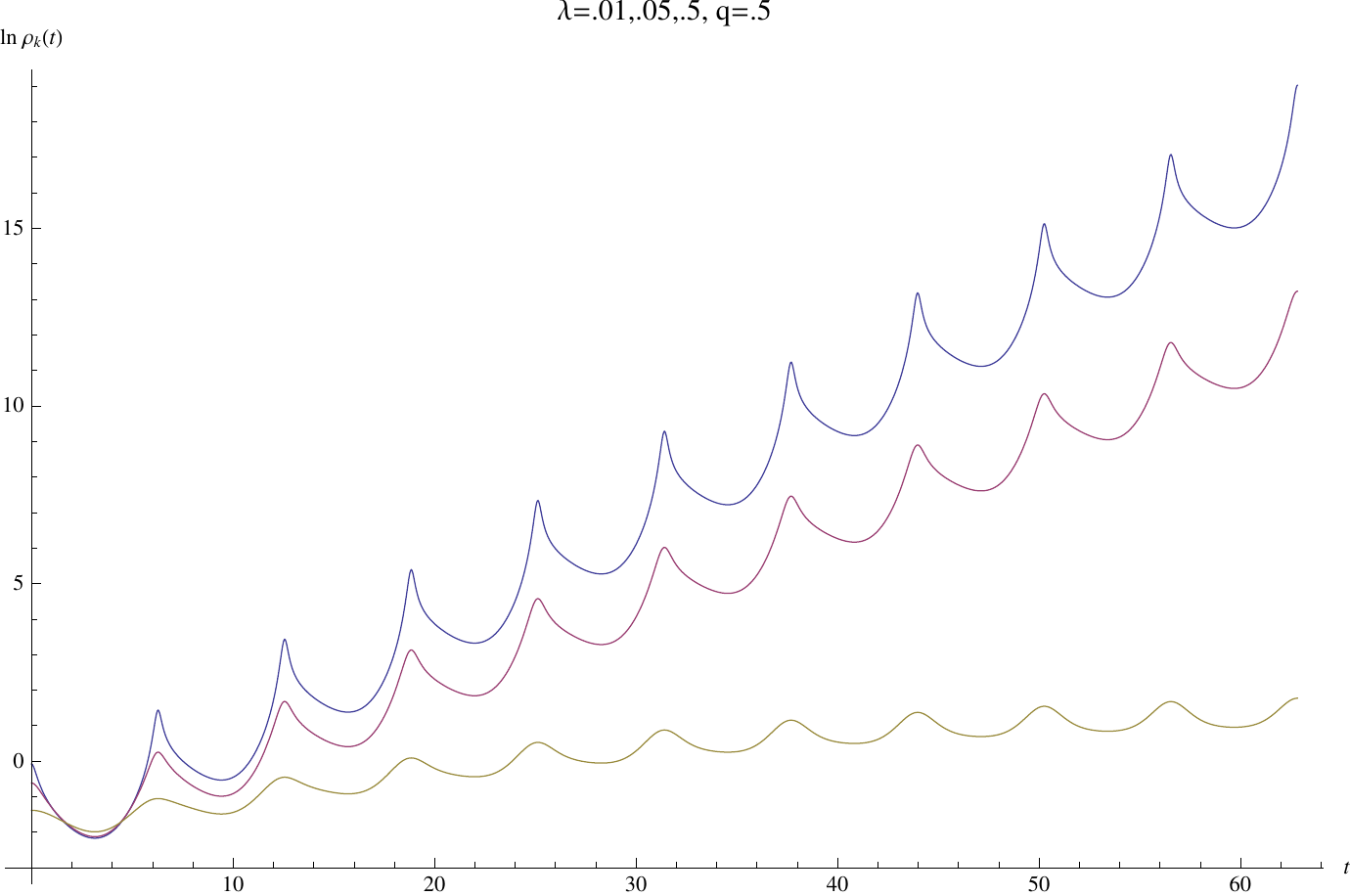}
\end{center}
\label{??}
\caption{The growth of energy density in unstable modes with $q=0.55$ and $\lambda = 0.5,~ 0.05,~ 0.01$.  Lower curves on the graph correspond to larger values of $\lambda$.}
\end{figure}

The parameter $q$ defined in Eq.~(\ref{q}) depends on $\lambda$, as well as on $L$ and the mode number $n^2$.  In Fig.~3 we show regions of instability for the independent parameters of the model, $\lambda$ and 
\beq
\kappa = (GL^2 \rho_{max})^{-1/2}.
\label{kappa}
\eeq
Each unstable band of Fig.~1 now splits into an infinite number of bands, corresponding to different values of $n^2$.  
We have included in Fig.~3 only modes with $n^2 \leq 10$, with higher values of $n^2$ indicated by lighter shades of grey.

\begin{figure}[t]
\begin{center}
\includegraphics[width=10cm]{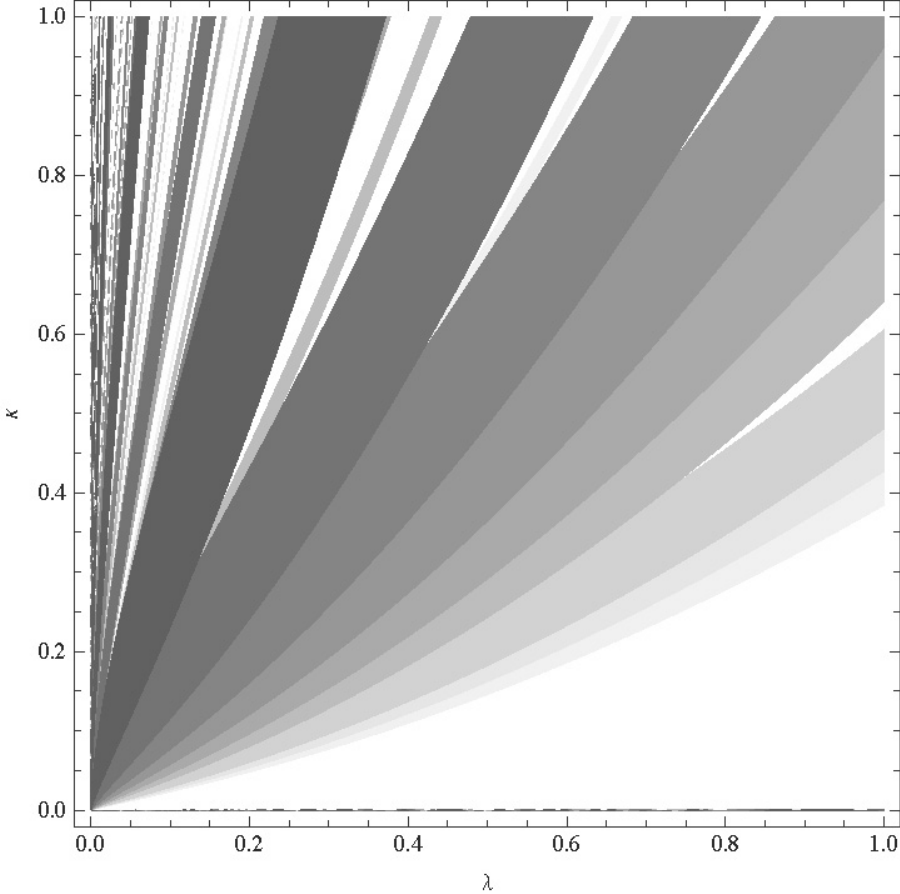}
\end{center}
\label{???}
\caption{Stability diagram in the parameter space of $\kappa$ and $\lambda$, with instability regions corresponding to different values of $n^2$ marked by different shades of grey.  Lighter shades correspond to larger values of $n^2$.}  
\end{figure}

It is interesting to compare our results with those of Graham et al \cite{Graham}, who studied the stability of their simple harmonic universe model.  In that model, the solution for $a(\tau)$ is inversely proportional to the oscillation frequency $\omega$, so $\omega$ drops out of the mode equation (\ref{phikeq}).  This leaves a single model parameter $\tilde\gamma = 3|\Lambda|/2\pi G\rho_0^2$, where $\Lambda <0$ is the cosmological constant and   
$\rho_0$ characterizes the contribution of matter with $w=-2/3$ to the total density.  The role of this parameter is similar to that of $\lambda$ in our model.
Graham et al find that for $\tilde\gamma \sim 1$, $a_{max}/a_{min}\sim 1$ and all modes are stable, while for $\tilde\gamma\ll 1$, $a_{max}/a_{min}\gg 1$ and there is a large number of unstable modes.  (An oscillating solution exists only for $\tilde\gamma < 1$.)  In contrast, our diagram in Fig. 3 shows a non-trivial pattern of stability and instability regions.  In particular, we find wide ranges of $\kappa$ where the model is unstable
when $\lambda\sim 1$ and where it is stable when $\lambda$ is very small.

\section{Tunneling to nothing}

Even if the oscillating LQC universe is stable with respect to classical perturbations, it may tunnel quantum mechanically from the bounce point at $V=V_{min}$ to a state of zero volume.   Here we shall estimate the semiclassical tunneling probability.

The application of standard semiclassical methods to this case is complicated by the fact that the field $\phi$ has a nonzero velocity at the bounce point, ${\dot\phi} (V_{min}) = p_\phi /V_{min}$ [see Eqs.~(\ref{dotp}-\ref{dotphi})].  This problem, however, is not difficult to resolve if we note that the variables $V$ and $\phi$ are essentially decoupled from one another.  The Hamiltonian (\ref{Heff}),(\ref{Hmatt}) is independent of $\phi$, so $p_{\phi}=(2C)^{1/2} ={\rm const}$, and the wave function can be factorized
\beq
\Psi(V,\phi) = e^{ip_\phi \phi} \psi(V).
\eeq
The problem then reduces to a one-dimensional tunneling problem with a single dynamical variable $V$, which is described by the action
\beq
S = \int_i^f dt \left(\frac{1}{4\pi G\gamma} \beta \dot{V} - N{\cal H} \right),
\label{reducedS}
\eeq
where
\beq
{\cal H} = -\rho_{max} V {\cos^2(\ell \beta)} + \frac{C}{V} +\Lambda V .
\label{HV}
\eeq

The semiclassical tunneling probability is given by
\beq
{\cal P} \sim e^{-2S_E} ,
\eeq
where $S_E$ is the Euclidean action of the instanton solution to the Euclidean equations of motion. The instanton can be obtained from the Lorentzian solution (\ref{at}),(\ref{beta}) by analytic continuation $t \to -i\tilde{t}$, $\beta \to i\tilde{\beta}$ to the classically forbidden range of $0<V<V_{min}$.  This gives
\beq
V({\tilde t})  =  L^3 (2\lambda)^{-1/2}
\left( -\cosh(\omega {\tilde t}) +1 + 2 \lambda \right)^{1/2} ,
\label{Vinst}
\eeq
\beq
\tilde{\beta}({\tilde{t}}) = - \left(\frac{8\pi G\gamma^2 \rho_{max}}{3}\right)^{1/2}
\tanh^{-1} \left[\sqrt{\frac{1+\lambda}{\lambda}}  \tanh \( \frac{\omega{\tilde t}}{2} \) \right] .
\eeq

The classically forbidden range extends from ${\tilde t}=0$, where ${\tilde\beta}=0$, to ${\tilde t} = {\tilde t}_f$, where $V=0$.  (${\tilde t}_f$ can be found from $\cosh (\omega {\tilde t}_f) = 1+2\lambda$.) Taking into account the Hamiltonian constraint ${\cal H}=0$, the instanton action is given by
\beq
S_E  = \frac{1}{4\pi G\gamma} \int_0^{{\tilde t}_f} d\tilde{t} \left| \tilde{\beta} \frac{dV}{d{\tilde t}} \right| 
= \frac{1}{4\pi G\gamma}  \left| \int_0^{V_{min}} dV \tilde{\beta}(V) \right| .
\label{SE}
\eeq

The form of $\tilde{\beta}(V)$ may be determined from the constraint,
\beq
\tilde{\beta}(V) = \ell^{-1} \cosh^{-1}\left[ \rho_{max}^{-1/2}\sqrt{\frac{C}{V^2}-|\Lambda|} \right] .
\eeq
Then, introducing a new variable $x=V/V_{min}$, 
the integral (\ref{SE}) can be rewritten as
\beq
S_E = \frac{V_{min}}{4\pi G\gamma \ell} \left| \int_0^{1} dx \cosh^{-1}\left({\frac{1+\lambda}{x^2}-\lambda} \right)^{1/2} \right| .
\eeq
It can be expressed as an elliptic integral, but this expression is not particularly illuminating and we do not present it here.  An interesting special case is the limit $\lambda\to 0$, when $V_{max}\gg V_{min}$.
In this limit we obtain
\beq
{\cal P}  \sim \exp \left(-\frac{\pi\sqrt{3}}{2} \frac{V_{min}}{\ell^3} \right) ,
\eeq
where we have used
\beq
\int_0^1 dx \cosh^{-1}\frac{1}{x} = \frac{\pi}{2}.
\eeq
The above semiclassical treatment indicates that when the universe reaches its minimum volume, there is a non-zero probability of tunneling to a singularity.  The tunneling is strongly suppressed when the minimum volume at the bounce $V_{min}$ is much larger than the Planck volume.

We note finally that our conclusions here are somewhat different from those of Ashtekar et al \cite{Ashtekar2010}, who studied quantum tunneling in the same LQC model with $\Lambda = 0$.  They found that the Euclidean action in the classically forbidden region between $V=0$ and $V=V_{min}$ is $S_E=0$, suggesting that the tunneling to $V=0$ is unsuppressed.  On the other hand, numerical calculations in Ref.~\cite{Ashtekar2010} indicate that the wave function is actually suppressed in this region,\footnote{We also mention a related result by Craig \cite{Craig} who showed analytically that eigenfunctions of the quantum evolution operator in LQC decay exponentially in the region between zero volume and the bounce.} and the authors interpret this as a breakdown of the semiclassical approximation.  In our view, the semiclassical approximation is accurate under the usual conditions (roughly, $S_E \gg 1$).  The reason for the discrepancy is that the Euclidean continuation in \cite{Ashtekar2010} was performed in the full action, including both $V$ and $\phi$ variables, while we considered a reduced action (\ref{reducedS}) with a single variable $V$. The latter appears to be the correct prescription in the presence of classical motion in $\phi$.\footnote{For a discussion of multi-dimensional tunneling in the presence of classical motion, see Ref.~\cite{Gregory}.}

\section{Summary and discussion}

\subsection{Particle production}

We investigated the stability of some oscillating universe solutions to the effective equations in Loop Quantum Cosmology.  These solutions have the topology of a 3-torus and contain matter in the form of a homogeneous massless scalar field and a cosmological constant $\Lambda < 0$.  We found that for some values of the parameters the model is stable with respect to small perturbations of the scalar field, indicating that oscillation of the scale factor does not result in particle production.  There are, however, substantial regions of the parameter space which are unstable.  

In a universe described by an unstable solution, there will be particle production.  The energy density of the created particles will grow exponentially with time and will eventually become comparable to that of the homogeneous field.  Our perturbative analysis cannot be extended beyond this point; here we shall attempt to give a qualitative outline of subsequent evolution.  For definiteness we shall assume that $\lambda\ll 1$, so that $V_{max}/V_{min}\gg 1$.

The additional matter component due to the created particles can be approximated as a perfect fluid with a radiation equation of state $w=1/3$, so the total energy density is
\beq
\rho = \Lambda + \frac{C}{V^2} + \frac{D(t)}{V^{4/3}} \equiv \Lambda + \rho_\phi + \rho_r,	
\label{rhoD}
\eeq
where $D(t)$ grows with time. As the universe contracts, the `radiation' energy density $\rho_r$ grows slower than 
$\rho_\phi$, so it will have little effect on the dynamics of the bounce at $a_{min}$, at least initially.  By the same token, $\rho_r$ decreases slower than $\rho_\phi$ during the expansion and will eventually exceed $\rho_\phi$ at large $a$.  As a result, it will take longer to reach the bounce at $a_{max}$ (where $\rho=0$), and the value of $a_{max}$ will increase.  The effect of this is similar to that of decreasing $\lambda$.

If the initial state at the onset of instability is in one of the narrow bands in Fig.~3, then any change of the parameters is likely to take us out of the resonance and to stabilize the oscillating universe.  On the other hand, if the initial state is in a broad instability band (say, at a relatively small $\kappa \sim 0.05 - 0.1$), then an effective decrease of $\lambda$ can make the solution even more unstable.  Eventually $\rho_r$ will become larger than $\rho_\phi$ even at small $a\sim a_{min}$.  The bounce will then occur at a larger value of $V_{min}$, which amounts to an effective increase in $L$ and a decrease in $\kappa$.  A look at the diagram in Fig.~3 suggests that if we start at relatively small values of $\kappa$ and $\lambda$ and move in the direction of still smaller $\kappa$ and $\lambda$, we may never leave the instability region.  This means that particle creation may be a runaway process, with both $a_{max}$ and $a_{min}$ becoming larger and larger and more and more particles being produced.  Whether or not this happens depends on the initial state, the other option being a stable oscillating universe.  This, however, is not the whole story.

\subsection{Quantum collapse}

Even if the oscillating universe is stable with respect to particle production, we found that every time it bounces at $a= a_{min}$, there is a nonzero probability for it to tunnel to zero size, $a=0$.  Classical solutions of GR become singular at $a=0$, and it is usually assumed that the spacetime terminates there.  Then a tunneling to $a=0$ would be the end of the universe.  

An alternative scenario is that the singularity at $a=0$ will eventually be resolved in the quantum theory of gravity.  It could be replaced, for example, by a Planck-size nonsingular nugget with $a\sim\ell$.  Then a tunneling $a_{min} \to \ell$ could be followed by the inverse process $\ell\to a_{min}$, and the oscillation would resume.  If this state of affairs can continue indefinitely, this could serve as a basis for a viable model of an emergent universe.  

However, if a tunneling $\ell\to a_{min}$ is possible, it seems likely that the parameters $C$ and $D$ characterizing the energy content of the resulting oscillating universe will generally be different from their initial values at the time of quantum collapse.  The cosmological constant $\Lambda$ may also be different if the particle physics model admits 
a landscape of vacua with different values of $\Lambda$.  The universe will then explore the entire parameter space and will eventually get into the regime of runaway particle production.  

The following argument suggests that quantum production of increasingly large numbers of particles may be a general feature of any model of a compact nonsingular universe.  The conserved energy-momentum `complex' $\Theta_\mu^\nu$ in GR can be expressed in terms of a `superpotential' $U_\mu^{\nu\sigma}=-U_\mu^{\sigma\nu}$ \cite{Landau},
\beq
\Theta_\mu^\nu = \partial_\sigma U_\mu^{\nu\sigma} ,
\eeq
\beq
\partial_\nu \Theta_\mu^\nu = 0.
\eeq
The total energy and momentum of a compact universe are then identically zero, 
\beq
P_\mu = \int d^3 x \partial_i U_\mu^{0i} = 0,
\eeq
with negative gravitational energy exactly compensating the energy of matter.  Hence, the production or destruction of particle-antiparticle pairs in such a universe are not forbidden by any conservation law.  As a rule of thumb in quantum mechanics, any process that is not strictly forbidden will occur with some probability.  Furthermore, an increase in the number of particles increases the entropy of the universe.  This indicates that a compact universe will evolve towards states with a larger number of particles.  In oscillating models of the kind we considered here, this is accompanied by an increase in the size of the universe.  Given an infinite time, the number of particles and the size of the universe would then grow without bound.  A quantitative analysis of these issues should await progress in the quantum theory of gravity.

\section*{Acknowledgements}

This work was supported in part by the National Science Foundation (grant PHY-1213888) and the Templeton Foundation.  We are grateful to Param Singh for very useful discussions.

\section*{Appendix}


In this Appendix we shall derive a stability criterion for solutions to the equation
\beq
\ddot{\phi}_k + 3 \frac{\dot{a}}{a} \dot{\phi}_k +\frac{k^2}{a^2}\phi_k = 0,
\label{eom}
\eeq
where the scale factor $a(t)$ is a periodic function with period $T$.  

According to Floquet's theorem, Eq.~(\ref{eom}) admits solutions of the form
\beq
\phi(t) = e^{i\alpha t} p(t),
\label{Floquet}
\eeq
where $p(t)$ is a periodic function with period $T$, and $\alpha$ is a constant defined by the boundary conditions.  It is apparent that when $\alpha$ is real, the solutions are oscillatory and therefore stable, whereas when $\alpha$ is complex, the solutions grow or decay and are unstable.

Note that if $\phi(t)$ is a solution of (\ref{eom}), then $\phi(t+T)$ is also a solution.  With the ansatz (\ref{Floquet}), we have
\beq
\phi(t+T) = \zeta \phi(t),
\label{rho}
\eeq
where $\zeta = e^{i\alpha T}$ is a constant.

To derive the stability criterion, we first define two solutions, $\phi_1(t)$ and $\phi_2(t)$, by the initial conditions
\begin{eqnarray}
\phi_1(0) = 1 \qquad  & & \qquad \phi_2(0) = 0 \nonumber \\
\dot{\phi}_1(0) = 0 \qquad & & \qquad  \dot{\phi}_2 (0) = 1.
\label{BC}
\end{eqnarray}

Any solution can be written as a linear combination of these two solutions: 
\beq
\phi(t) = c_1 \phi_1(t) + c_2 \phi_2(t).
\label{c1c2}
\eeq
In particular, the solutions $\phi_1(t+T)$ and $\phi_2(t+T)$ are
\begin{eqnarray}
\phi_1(t+T) &=& \phi_1(T) \phi_1(t) + \phi_1'(T) \phi_2(t) \nonumber \\
\phi_2(t+T) &=& \phi_2(T) \phi_1(t) + \phi_2'(T) \phi_2(t).
\label{phiT}
\end{eqnarray}

For any solution of the form (\ref{rho}), we obtain, using Eqs.~(\ref{c1c2}), (\ref{phiT}), the following set of linear equations for $c_1$ and $c_2$:
\begin{eqnarray}
(\phi_1(T) -\zeta) c_1 + \phi_2(T) c_2 &=& 0 \\
\phi_1'(T) c_1 + (\phi_2'(T) -\zeta) c_2 &=& 0
\end{eqnarray}

Nonzero solutions to this set of equations exist when
\begin{eqnarray}
\begin{vmatrix}
(\phi_1(T) -\zeta) &  \phi_2(T) \\
\phi_1'(T) & (\phi_2'(T) -\zeta)
\end{vmatrix} =  \zeta^2 - (\phi_1(T) + \phi_2'(T)) \zeta + W(\phi_1(T),\phi_2(T)) = 0
\label{characteristic}
\end{eqnarray}
where $W(t) = W(\phi_1,(t),\phi_2(t))$ is the Wronskian.  From Eq.~(\ref{eom}), $W(t) a^3(t) = \text{const}$, so that $W(T) = W(0)$.  With the boundary conditions in Eq. (\ref{BC}), we have $W(0) =1$, so that Eq.~(\ref{characteristic}) becomes 
\beq
 \zeta^2 - (\phi_1(T) + \phi_2'(T)) \zeta + 1 = 0.
\eeq

The product of the two roots of this equation is equal to 1; hence the roots can be represented as $\zeta_{1,2} = \exp(\pm i\alpha T)$.  The sum of the roots is
\beq
\zeta_1 +\zeta_2 =2\cos (\alpha T) =  \phi_1(T) + \phi_2'(T) \equiv b.
\eeq
If $b > 2$, then the roots are real, while if $b < 2$, the roots are complex and conjugate to one another.



As we indicated above, the solutions are stable for real $\alpha$ and unstable for complex $\alpha$.  This corresponds to the condition
\beq
|\phi_1(T) + \phi_2'(T)| < 2
\label{stab}
\eeq
for stable solutions, and 
\beq
|\phi_1(T) + \phi_2'(T)| > 2
\label{instab}
\eeq
for unstable solutions.\footnote{ The case where $|\phi_1(T) + \phi_2'(T)| = 2$, corresponding to the boundary between the stable and unstable regions, is further analyzed in \cite{Magnus}.}  In order to check the mode stability in our model, we integrate Eq.~(\ref{eom}) with $a(t)$ from Eq.~(\ref{at'}), using the initial conditions in (\ref{BC}).  Then we check whether the solutions at $t=T$ satisfy (\ref{stab}) or (\ref{instab}).

The stability diagram in Fig.~3 was produced by sampling the parameter space $0<\lambda,\kappa<1$ at logarithmic intervals, so there are many more points at small values of the parameters than at $\lambda,\kappa\sim 1$.  This ensures that we have high resolution in regions where it is required.

\end{document}